\documentstyle[NATO,numreferences]{crckapb}
\begin{opening}

\title{POINT-CONTACT SPECTROSCOPY OF SUPERCONDUCTORS}
\author{I.K. Yanson}
\institute{B. Verkin Institute for Low Temperature Physics and Engineering\\
National Academy of Sciences of Ukraine, 310164 Kharkiv, Ukraine}
\end{opening}
\runningtitle{POINT-CONTACT SPECTROSCOPY}
\begin{document}

keywords: point-contact spectroscopy, electron-phonon interaction, S-c-S and 
S-c-N contacts, excess current

\section{Introduction}

In this review we describe the current-voltage spectroscopy of
superconductors which, in contrast to the tunneling spectroscopy,
considers 
contacts with direct conductivity, without any barrier. These contacts are
usually called point contacts as their lateral
dimension should be smaller than some microscopic lengths such as
the coherence
length in superconductors and the inelastic mean free path of
conduction electrons in
normal metals. Some of 3-dimensional metallic constrictions are
schematically depicted in Fig.1. A shorted thin-film tunneling junction (a)
was historically the first structure in which the point-contact
spectroscopy was
discovered \cite{Yan1974}. A much more sophisticated thin-film structure is
displayed in (b) where the tiny controlled orifice was made in the insulating
diaphragm of Si$_3$N$_4$ by means of ion-beam nanolithography \cite{Ralls}.
Point-contacts with bulk electrodes are shown in Fig1 (c-e) for different
geometries. The simplest are the pressure-type contacts in which two sharp edges
(c) \cite{Chubov} or a needle and an anvil (d) \cite{Jansen} are
forced to 
touch  each other gently. Most point-contact spectroscopic studies
were made just with these very simple experimental devices. The mechanically
controlled break-junction (MCB) (Fig.1 (e)) gives an experimentalist
new advantages \cite{Muller}. A small rode of a metal 1 is glued
by the Staycast varnish 3 to the bending beam 4. The latter can be bent by
a piezo driver 5 so that the notch 2 at the centre of the
sample becomes stretched and the cross-section area of the constriction is
gradually decreased. This device allows us to study the contact properties as a
function of constriction size.

In this review we do not concern the so-called Andreev-reflection
spectroscopy of superconducting energy gap. The reader can find this in
Refs. \cite{BTK,AVZ,Zaitsev} and later
extensions \cite{Dynes,Plecenik}. Our focus will be on the
above-gap-energies where one could expect nonlinearities caused by the
electron-boson interactions responsible for the formation of Cooper pairs.
The
problem unsolved yet is that in standard superconductors (lead,
tin, Nb$_3$Sn and many others) the above-gap nonlinearities give the
evidence about the electron-phonon-interaction mechanism of Cooper pairing 
\cite{Rowell}, whilst the tunnel junctions of unconventional
superconductors (high-T$_c$'s, heavy
fermions, organic superconductors, etc.) do not show any noticeable
nonlinearities for energies larger than the superconductor energy gap \cite
{Fisher,Ekino} (albeit see Ref.\cite{UPd2Al3}). On the contrary, the point
contacts do manifest  strong nonlinearities at above-gap energies which in
many cases correlate with the boson density of states known, for instance, from
neutron study \cite{LaSrCuO,Rev4}.

We thus think that the study of nonlinearities of point contacts is the
alternative way of investigating the spectral functions of the
electron-boson interaction which in many cases is more productive and
simpler than tunneling.

\section{Point-contact spectroscopy in the normal state}

First, we consider the point-contact spectroscopy (PCS) of
electron-boson interaction (EBI) in the normal state \cite{Rev1}. Many
superconductors can be driven into the normal state by magnetic field, and
the PCS theory is more fundamental and transparent in this case. As a
model for 3D constrictions, we consider a circular channel of length $L$ and
diameter $d$ connecting two bulk metallic half-spaces (Fig.2). If $L\ll d$,
we have an orifice in nontransparent infinitely thin partition, while for
the opposite case ($L\gg d$) a one-dimensional channel appears. Since
contact sizes $L,d$ are much less than the mean free path of conduction
electrons {\bf e}$^{-}$ the latter move through the constriction
ballistically on applying the voltage bias $eV$. The resistance of such a
ballistic contact was first derived by Sharvin \cite{Sharvin} and is equal to

\[
R_{Sh}=R_q\left( \frac{16}{d^2k_F^2}\right) ;\;\;\;R_q=\frac{\pi
\hbar }{e^2}\simeq 12.9 \: \mathtt{ k}\Omega 
\]

Suppose that there are some seldom elastic scatterings off static defects or
impurities ($\star $) or inelastic processes with emission phonons and other
quasiparticles shown in Fig.2 as a wavy line with momenta $q$ . Let us call
this regime quasi-ballistic. The nonequilibrium distribution function in the
momentum space consists, like in the ballistic regime, of two spherical
segments which are displaced in energy by $eV$ and which have the form of
two semispheres at the central cross-section of the constriction \cite{KOS}.
At low temperatures we can neglect the smearing of the Fermi edge. Then, the
energy conservation sets a sharp threshold $\hbar \omega \leq eV$ on the
quasiparticles which can be emitted by the inelastic transition of conduction
electron with momenta $p$ to $p^{^{\prime }}$. This backscattering processes
create opposite current which depends strongly on the bias because of the
energy dependence of the quasiparticle density of states and matrix element of
EBI. This current is the basis of PCS. To be complete, we must note that in some cases 
the elastic scattering can also depend on the energy like, for
instance, the scattering off the paramagnetic impurities (in Fig.2 these are
shown with $\nearrow $) due to the Kondo effect or to two-level systems.

In the quasi-ballistic regime $\left( l>\max \{d,L\}\right) $ the
differential resistance of the point contact is equal to

\begin{eqnarray*}
R(eV) &=&R_{Sh}+\Delta R(eV);\;\;\mbox{ where }\;\;\Delta R(eV)/R_{Sh}=\left( 8/6\pi
\right) [d/l_i(eV)]. \\
\end{eqnarray*}

For the second derivative of the current-voltage characteristic, which is
usually called the point-contact spectrum, we obtain at $T\simeq 0$ a simple
relation \cite{KOS}:

\begin{equation}
\frac{d\ln R(eV)}{d(eV)}=\frac 83\frac{ed}{\hbar v_F}g_{PC}(eV)_{}
\label{spectrum}
\end{equation}

where g$_{PC}(eV)$ is the transport EBI function which in many cases appears
to be the electron-phonon-interaction (EPI) spectral function similar to
the Eliashberg EPI function \cite{KOS}. The point-contact EPI function can
be expressed as $g_{PC}(\omega )=\alpha _{PC}^2(\omega )F(\omega )$, where $%
\alpha _{PC}^2(\omega )$ is the averaged electron-phonon matrix element with
kinematic restrictions imposed by the point contact geometry and $F(\omega )$ is
the phonon density of states. Thus, we see that PCS allows direct
recording of the transport version of EBI (EPI) spectral function which is
responsible for the formation of Cooper pairs in superconductors. We show
here only one example for tin although many pure metals, alloys and
compounds have already been studied by means of
PCS \cite{Jan,Rev2,Rev3,Duif,Rev1,Rev4}.
In Fig.3 (upper panel) the directly recorded PC spectrum of tin is shown by
a solid line which is compared with the EPI spectral function (dots) reconstructed
from the Eliashberg integral equations by means of the Rowell-McMillan procedure
from the superconducting tunneling experiment. Both curve are smeared
by the PCS resolution which is shown by the horizontal segment on the
graph and which is equal to 
\[
\delta (eV)=\sqrt{(5.44k_FT)^2+(1.72eV_1)^2}. 
\]
The PCS resolution depends on the temperature $T$ and modulation voltage $%
V_1 $ which is used for recording the second derivative of current-voltage
characteristic with a lock-in technique.

For symmetrical heterocontacts (point contact with dissimilar electrodes), the PC
spectrum at $T\simeq 0$ takes the form:

\begin{equation}
\frac{d\ln R(eV)}{d(eV)}=\frac 43\frac{ed}{\hbar }\left\{ \frac{%
g_{PC}^{(1)}(eV)}{v_F^{(1)}}_{}+\frac{g_{PC}^{(2)}(eV)}{v_F^{(2)}}\right\}
\label{hetero}
\end{equation}
where the superscript $(i)$ refers to particular metal. This formula is used for
a S-c-N contact.

The heavy elastic scattering around the contact is not necessarily
detrimental to PCS as long as the energy relaxation length remains much
larger than the contact size: 
\[
d\ll \sqrt{l_il_e}. 
\]
In the case of diffusive movement of charge carriers through the contact $%
\left( l_e\ll d\right)$, the contact size $d$ in formula (\ref{spectrum})
should be replaced by the elastic mean free path $l_e.$ The interpolation
curve of the spectrum intensity is shown on the lower panel of Fig.3 \cite
{KulYan}. It can be used for estimation of $l_e$ inside the contact by
measuring the intensity of the PC spectrum. This is important since during making a
contact one could introduced a lot of defects into the contact region.

\section{Excess current: dependence on purity}

If one or two of the electrodes of a point contact become superconducting,
the excess current emerges, which is superimposed on top of the normal state
current. The mechanism is called the Andreev transformation and is as follows:
a quasiparticle coming through the constriction has a certain
probability to find another electron to form a Cooper pair in the
superconducting electrode. This process leads to reflection of
quasiparticles which have  opposite signs of charge and velocity in the normal metal
(in case of S-c-N junction) but the same excitation energy \cite
{Andreev}. For biases larger than  gap energy this current is
constant if strong-coupling, inelastic and nonequilibrium
processes are disregarded. Below we consider all of them, but now let us focus on
the undisturbed excess current which spreads not too far from the gap region. In
The IVC of S-c-S tin junction is plotted in Fig.4 (upper panel) \cite{parallel}.
The normal-state zero-bias resistance is compensated by a bridge circuit,
hence the deflection of the real IVC in the normal state from the Ohmic behaviour is
shown by curve 2. Curve 1 presents the superconducting state of the same
contact ($H=0$). The critical (Josephson) current is clearly visible at $V=0$,
and a small dip is seen at $eV=2\Delta $. This is because the small barrier
appears
at the interface between the electrodes due to introduced inhomogeneities
while making the contact. The difference between curves 1 and 2 presents the
excess current for biases greater than $2\Delta $ which is seen to be
approximately constant up to the energies near 5 meV. We shall discuss the
nonlinearities of excess current

\begin{equation}
I_{exc}(V)=I_{exc,0}+\delta I_{exc}(V)  \label{excess}
\end{equation}
later on and now concentrate on the dependence of $I_{exc,0}$ on purity of the
metal. Here we have to consider the purity inside the constriction region
which could be quite different from that of the bulk. As a criterion of purity, we
may take the maximum intensity of PC spectra which depends on the elastic
mean free path averaged over the contact region, as shown by the curve in
Fig.3 (lower panel). The dependence of $I_{exc,0}$ in  $\Delta/eR_0$ units
is shown in the lower panel of Fig.4 as a function of $g_{PC}^{\max }$,
where $g_{PC}^{\max }$ is the value of the PC spectrum at the bias around 15 meV
(see Fig.3 (upper panel)). $I_{exc,0}$ starts approximately from $1.47$ for dirty
contacts (i. e. for small $g_{PC}^{\max }$), as was predicted theoretically by
Artemenko, Volkov and Zaitsev (AVZ) \cite{AVZ}, and rises with $g_{PC}^{\max
}$ up to the value predicted by Zaitsev and BTK \cite{Zaitsev,BTK} for clean
contact\footnote{%
For a S-c-N contact the excess current is equal to half of the theoretical
values shown in Fig.4 (lower panel).}. In Fig.4 (lower panel) we show
this ultimate value at $g_{PC}^{\max }=0.41$ taken from Ref.\cite{Atlas},
as a dot in a circle. From the graph $g_{PC}/g_{PC}^{\max }(l_e/d)$ of Fig.3
(lower panel) we find that $l_e/d\simeq 0.5$ is quite sufficient to transform a
clean S-c-S junction into a dirty one. On the other hand, the BTK theory
dealing with {\it clean} junctions gives the AVZ-dirty value for $%
I_{exc,0}$ at the barrier parameter Z$\simeq 0.5$ and, what is the most strange,
the BTK-fitting leads to approximately the same Z for large diversity of
materials, including high-T$_c$ , heavy fermions and conventional
superconductors. One might suspect that application of the clean BTK
procedure to dirty junctions gives the same Z$\simeq 0.5$ value
irrespective of the kind of material.

The parallel study of excess current and intensity of PC spectra allows one
to conclude whether the impurities are homogeneously distributed or they are
located as a thin barrier at the interface between the electrodes. Indeed,
the relative intensity of PC spectra does not saturate for $l_e/d\ll 1$
being proportional to $l_e/d$ , while the excess current does. The same is
true when we consider the Z-parameter, instead of the excess current,  while
fitting the IVC by the BTK theory.

\section{Elastic contribution to excess current}

Further on we consider S-c-N junctions which are simpler to interpret due to
the absence of the Josephson effect. To be more definite, we restrict ourselves to
the electron-phonon interaction. According to Refs.\cite{Khlus,Omel}, one can
write the IVC in the superconducting state as 

\begin{equation}
I(V)=V/R_0-\delta I_{ph}^N(V)+I_{exc}(V)  \label{nonlinear}
\end{equation}
where $\delta I_{ph}^N(V)$ is the backscattering current in the normal state
whose second derivative is given by Eq.(\ref{hetero}) and $I_{exc}(V)$ is
equal to Eq.(\ref{excess}). The order of magnitude of $\delta I_{ph}^N(V)$
is $I(V)\times \left( d/l_{in}\right) $. The voltage dependent part of $%
\delta I_{exc}(V)$ can be decomposed in elastic and inelastic parts:

\[
\delta I_{exc}(V)=\delta I_{exc}^{el}(V)+\delta I_{exc}^{in}(V) 
\]

Correspondingly, the order of magnitude of the $\delta I_{exc}^{el}(V)$ and $%
\delta I_{exc}^{in}(V)$ terms amounts to

\begin{equation}
I_{exc,0}\times (\Delta /\hbar \omega _{ph})\;\;\mbox{ and }\;\;I_{exc,0}\times
(d/l_{in})\;\;\mbox{ where }\;\;I_{exc,0}\simeq \Delta /eR_0.  \label{el-in-exc}
\end{equation}

Let us consider the elastic component which is essential for strong
coupling superconductors when $\Delta $ is not much less than $\omega _{ph}$%
. Fig.5 shows the experimental second derivatives of IVC for lead in
the superconducting state of Pb-Ru and Pb-Os point contacts (curves 2 and 4,
respectively). Normal metals Ru and Os were chosen since their phonon
density of states can be neglected in the energy range where Pb has the highest
intensity \cite{Yan-el}. The normal state EPI PC spectrum for contact 2 is
shown as dotted curve 3 on the same ordinate scale like that for the
superconducting state (curve 2). It is clearly seen that in the
superconducting state the phonon peaks are shifted to the higher energies
roughly by $\Delta $(Pb)$\simeq $1.3 mV. Moreover, the intensity of the peaks is
larger than that in the normal state. Importantly, the phonon peaks for
noticeably dirtier contact (curve 4) are not much smaller, as could be
expected from the measurements of EPI-PC-spectrum intensity in the normal
state (not shown). These properties qualitatively contradict the theoretical predictions
\cite{Khlus,Omel} if only the inelastic processes are taken
into account.

Let us compare quantitatively the spectra for a clean contact (curve 2) with
the predictions of Ref.\cite{Omel-el} which takes into account the
strong-coupling effect of the superconducting energy gap dependence on 
energy. For the ballistic S-c-N junction with a strong-coupling superconductor the
theory gives the expression of the first derivative of IVC:

\[
\left\{ \frac{dI}{dV}(eV)\right\} _{S-c-N}=\frac 1{R_0}\left\{ \frac{\Delta
(eV)}{eV+\left[ \left( eV\right) ^2-\Delta ^2(eV)\right] ^{1/2}}\right\} 
\]
for the total differential conductance at $T=0$. The derivative of this
curve is displayed as curve 1 in Fig.5 where we used the strong-coupling gap
versus energy dependence for lead determined by tunneling spectroscopy \cite{Rowell2}%
. The smooth weak-coupling BCS background (corresponding to $\Delta
(eV)=const$) is shown as a dashed line superimposed on curve 1.
The experimental characteristic (curve 2) agrees reasonably well with 
theoretical predictions (curve 1) both in shape and in amplitude. The latter
can be estimated by the deflection from the smooth background. The total
order of magnitude of the elastic correction to the conductance 
amounts to $\left[ \Delta (eV)/\hbar \omega _{ph}\right] ^2$ for $%
\Delta (eV)\ll \hbar \omega _{ph}$, just as in the tunneling spectroscopy
of superconductor. Since the elastic term of excess current $\delta
I_{exc}^{el}(V)$ does not depend on the contact diameter $d$ (\ref{el-in-exc}%
), it can be higher than the inelastic contribution $\delta I_{exc}^{in}(V)$%
. This situation occursjust for lead. Below we consider another case which
holds for weak-coupling superconducting tin, where the inelastic term
prevails. We conclude this section with the notion that the Eliashberg EPI
spectral function can be extracted from the superconducting point-contact
characteristic in the same way as in tunneling spectroscopy provided that one
should be sure that the inelastic terms are negligible.

\section{Inelastic processes in excess current}

For a weak-coupling superconductor ($\Delta (eV)\ll \hbar \omega _{ph}$)
the elastic contribution to the excess current is negligible. Consider the Sn-Cu
S-c-N point contact which is shown schematically as an inset in Fig.6 \cite
{Yan-inel}. First of all, in the normal state Sn and Cu have
approximately the same electronic parameters ($v_F$) and overlapped energy
regions of the phonon density of states. Their PC\ EPI spectra are shown as
curves 5 and 6 for Sn and Cu, respectively. In a symmetrical heterocontact
their spectral functions enter almost equally according to the formula (\ref
{hetero}) $g_{PC}$(Sn-Cu)$\simeq \left( 1/2\right) \left[ g_{PC}\mbox{(Sn)}%
+g_{PC}\mbox{(Cu)}\right] $ which is shown as dotted curve 4. Accordingly,
the measured PC spectrum in the normal state of the heterocontact looks like
curve 3 which (with a small smooth background subtracted) leads to 
solid curve 4. The coincidence between dotted and solid curves 4 not only in
shape but also in amplitude is fairly well. In the superconducting state of
the same contact the excess current appears, which is shown by curve 1 in
Fig.6 as the difference between the total current and the Ohmic term $V/R_0$ in
superconducting and normal states, respectively. Evidently, the excess
current is not constant. We disregard the large hysteresis-like drop at high
voltage (larger than the phonon energy band) due to heating and concentrate
our attention on the smooth decrease along the phonon energy range. The second derivative
(PC spectrum) of IVC in the superconducting state is shown as curve 2. Apart
from the smooth background which steeply rises when the voltage approaches
the energy gap ($\Delta $(Sn)$\simeq 0.6$ meV) from above, one sees the curve
which almost reproduces the features seen in the normal-state spectrum
(curve 3). We thus conclude that the main nonlinearities in the
superconducting state come from the backscattering current the same as in
the normal state. Yet, there are small peculiarities such as the
small shift of the maxima to the lower energy (shown by the vertical dashed
line in Fig.6) and a slight broadening of peaks. These peculiarities are
predicted by  theory \cite{Khlus}. The unexpected shift to the lower
energy by the superconducting energy gap is due to electron-hole relaxation
as a result of the Andreev reflection from the N-S boundary. The inelastic
processes with emission of phonons occur when an electron quasiparticle with
energy of the order of $eV$ relaxes to the hole state appearing due to
Andreev reflection of another electron with energy of the order of $\Delta $%
. For an infinitely narrow peak in the EPI function the theory leads to the
maximum shifted down by $\Delta $ with the width of the order of $\Delta $ as
well. Hence, in the clean S-c-N contact based on a weak-coupling
superconductor the main contribution to formula (\ref{nonlinear}) comes
from the normal term $\delta I_{ph}^N(V)$ with slight modifications due to $%
\delta I_{exc}^{in}(V)$ term.

\section{Nonequilibrium phenomena}

As soon as the diameter of the contact becomes comparable with the
superconducting coherence length $\xi _0$, the nonequilibrium phenomena may
occur which lead to suppression of the gap by i) nonequilibrium populations of normal
electrons and phonons with energies greater than $2\Delta $, ii) entering the
magnetic flux vortices generated by the current through the contact, or iii)
simply by heating of the contact region.

In Fig.7 the PC spectra of Ta-Cu contact are shown both in the normal (curve 1)
and superconducting (curve 2) states \cite{Yan-noneq}. Since the Fermi
velocity of Ta is noticeably less and the strength of EPI in Ta is
essentially higher than for Cu, only the EPI spectrum of Ta is seen in the
experiment: $g_{PC}^{(Ta-Cu)}\simeq \frac 12g_{PC}^{(Ta)}$ (see Eq.(\ref
{hetero}). Although the relatively high resistance of contact ($R_0=80$ $%
\Omega $) implies a small size $(d\simeq 78$ \AA ), the ultimate (at $%
eV>\hbar \omega _{ph}$) electron-phonon mean free path is also small ($%
l_{in}\sim 120$ \AA ) and the reabsorption of nonequilibrium phonons may
occur. Also the superconducting coherence length in Ta ($\xi _0=90$ nm) is
smaller than in Sn ($\xi _0=270$ nm), especially for the dirty Ta-contact region.
This is more favourable for nonequilibrium transitions in the superconducting
state and assists, by decreasing $H_{c1}$, the current-generated vortices
to enter the contact area.
All this makes the EPI spectrum in the superconducting state of Ta quite
different from what is observed in Sn.
There is no shift of the phonon peaks by $\Delta $ to lower biases. Instead, the
lower is the contact resistance (the smaller is the size), the more fixed are
the transverse (T) and
longitudinal (L) phonon peaks at energies of 11.3 and 18 meV
with very small scattering of $\pm 0.1$ meV. The peak of T-phonon (see curve 2 in
Fig.7) grows narrower instead of broadening and with increase of the contact
size all the peaks become sharpened (Fig.8). This contrasts to the behaviour of
phonon peaks in clean Ta point contacts in the normal state where one
observes the spreadingss of energy positions in the limit of $\pm 1$ meV
probably due to the random anisotropy of the microcrystal orientation with
respect to the contact axis.
The explanation is as following. The nonequilibrium phonons slowly diffuse
from the dirty contact region effectively averaging their direction in the
momentum space. The phonon peaks at reproducible fixed positions appear due to
the slow phonon group velocity (corresponding to the maximum of the averaged
phonon density of state) which are accumulated in the contact region and
thus effectively depresses the superconducting energy gap and sxcess curent.
In addition to the common energy-gap feature at
$eV\simeq 0.6$ meV, a new pecularity appears at $V_1$ (at about 5$\div 7$
meV in Fig.7), which has strong intensity (note that the ordinate of this
fragment has a factor of 0.01!), and whose voltage satisfies the relation $%
V_1^2/R_0=const$. The latter means that either the power input by the
current or the magnetic field generated at the contact achieve the threshold
value. The threshold value increases with temperature and external magnetic
field which is just opposite to the behaviour expected for destruction of
superconductivity by simple heating or critical current conditions. We infer
that at larger biases the superconductor turns into the nonequilibrium state
and the increase in threshold value is explained by enhanced relaxation of
nonequilibrium quasiparticles which demand a stronger injection to fulfil
the threshold conditions. In this state the phonon features remain at the
proper energies and are seen as a sharp singularity on IVC and its derivatives (Fig.8),
whilst in the normal state only the smooth background without any spectral
features is observed. Although there is no theory explaining quantitatively
these phenomena yet, the experimental extension of the superconducting
PCS to the materials with a short inelastic mean free path and coherence
length seems to be very encouraging in view of studies on  exotic
superconductors, such as high-T$_c$, heavy fermions and organic compounds.

\section{Concluding remarks}

We have briefly described the state of the art of point-contact spectroscopy in
the superconducting state. It allows one to penetrate into the mechanism of
electron-quasiparticle interaction which may mediate the Cooper pairing.
Contrary to tunneling spectroscopy, the point-contact spectroscopy needs not
any barrier which inevitably interrupts the homogeneity of the material studied
and often hampers  investigation of many complicated compounds. At the expense of
this, PCS results in principle to the highly nonequilibrium state which in case of
superconductors leads to many complications. We hope that with development
of new technique of producing well controlled constrictions of nanoscale
size and elaborating a new theoretical approach the PCS will conquer more
and more supporters and will become in the future as important as the
tunneling spectroscopy.

The author wishes to acknowledge the financial support of the International
Science Foundation (Soros Foundation).

\newpage\ 
\begin{figure}[tbp]
\caption[]{
The three-dimensional constrictions made by different technique: (a) shorted
thin-film tunneling structure \cite{Yan1974}; (b) thin-film point contact
made by nanolithography \cite{Ralls}; (c) contacted sharp edges of bulk
metals; arrows show the movement of electrode to choose the proper spot for
measuring \cite{Chubov}; (d) ''spear-anvil'' geometry of the bulk electrodes 
\cite{Jansen}; (e) mechanically controllable break-junction \cite{Muller};
arrow shows the movement of piezo drive 5 in order to bend the substrate 4.
}
\end{figure}

\begin{figure}[tbp]
\caption[]{
Schematic illustration of quasi-ballistic electron flow. Two metallic half
spaces are connected by a circular orifice of diameter $d$ in a partition of
thickness $L$. Voltage bias $eV$ is applied between them. Most of the
electrons flow ballistically through an orifice but some of them experience
rare collisions off static (denoted by stars) and dynamic (denoted by wavy
lines) scatterers. Small arrows superimposed on stars denote the impurities
with magnetic moments. The dashed circle approximately denotes the area
where the nonlinear information to IVC comes from. In the lower part, the
nonequilibrium distribution function in the momentum space at the central
cross section is drawn. Inelastic backscattering from {\bf p} to {\bf %
p'} with emission of quasiparticle (phonon) with momentum {\bf q} is
shown. These processes have a sharp threshold $\hbar \omega \leq eV$ at
low temperatures.
}
\end{figure}

\begin{figure}[tbp]
\caption[]{
(upper panel): Comparison of electron-phonon-interaction spectral functions
determined by the superconducting tunneling reconstruction technique \cite
{Rowell2} (dotted curve) and directly recorded by the point-contact
spectroscopy \cite{parallel} (solid curve). The small horizontal bar denotes
the resolution for given parameters.

(lower panel): Interpolation curve for spectral intensity between the
ballistic limit $(l_e\gg d)$ and diffusive regime $(l_e\ll d)$ \cite{KulYan}.
}
\end{figure}

\begin{figure}[tbp]
\caption[]{
The parallel study of S-c-S contacts in superconducting ($H=0$) and normal $%
(H=0.9$ kOe$)$ states of tin \cite{parallel}. (Upper panel): the excess
current as a difference between curves 1 and 2 is plotted versus the voltage
bias. Contact resistance $R_0=2.55$ $\Omega .$ (Lower panel): Dependence of
excess current in units of $\Delta /eR_0$ on the value of spectral EPI
function (determined in the normal state) at the bias of $\simeq $15 meV.
Two horizontal straight lines determine the dirty and clean limits according
to Refs.\cite{AVZ,Zaitsev}.
}
\end{figure}

\begin{figure}[tbp]
\caption[]{
Phonon features on the PC spectra of S-c-N contacts for the strong-coupling
superconductor. Curves 2 and 4 correspond to the clean Pb-Ru contact  and
the dirty Pb-Os one, respectively. They are compared with 
theoretically predicted curve 1 which in turn should be compared with the
weak-coupling BCS limit shown with the dashed curve. Dotted curve 3 is
the EPI PC spectra of junction 2 with the same ordinate scale for both
cases. The curves are shifted vertically for clarity. The letters T and L
with arrows mark the positions of transverse and longitudinal phonon peaks
in lead, respectively.
}
\end{figure}

\begin{figure}[tbp]
\caption[]{
Inelastic phonon peculiarities for weak coupling superconductor in Sn-Cu
S-c-N junction (see inset). The upper curve shows the dependence of excess
current on voltage bias for contact with $R_0=8.8$ $\Omega $ with $d=10$ nm.
The second derivative of IVC in superconducting state is shown as curve 2.
Curve 3 gives the EPI spectrum for the same contact in the normal state.
It results in the EPI spectral function shown by solid line in curves 4
which should be compared with the calculated one plotted as a dotted curve.
Curves 5 and 6 display the PC EPI spectra of Sn and Cu homocontacts,
respectively. The curves are displaced vertically for convenience. Note the
different zeros along the ordinate axis for each of them.
}
\end{figure}

\begin{figure}[tbp]
\caption[]{
Nonequilibrium phenomena in PC spectrum of Ta-contact in the superconducting
state. The Ta-Cu contact is made by the ''edge-edge'' geometry as shown in
the inset. It has $R_0=80$ $\Omega $ and $d=78$ \AA . In the normal state ($%
H=3$ kOe) the spectrum is directly recorded as curve 1. Its shape 
practically coincides with the EPI spectral function calculated from the
superconducting tunneling \cite{Wolf}. In the superconducting state (curve 2)
anomalous features are seen. The transverse phonon peak at 11.5 meV
sharpens, and a part of the spectrum becomes negative. There appears a
strong nonlinearity at $\simeq $6 meV which looks like a peak on the
differential resistance curve and a sharp decrease of excess current on the
IVC.
}
\end{figure}

\begin{figure}[tbp]
\caption[]{
The evolution of nonequilibrium peculiarities with decreasing contact
resistance (increasing size). Curve 1: contact Ta-Cu, $R_0=26.5$ $\Omega $, $%
d=13.6$ nm. Curves 2 and 3: contacts Ta-Au with $R_0=19$ $\Omega $ $(d=16$
nm) and $R_0=0.76$ $\Omega $ ($d=80.3$ nm), respectively.
}
\end{figure}

\end{document}